\documentclass[fleqn,usenatbib,letters]{mnras}

\usepackage{newtxtext,newtxmath}
\usepackage[T1]{fontenc}

\usepackage{graphicx}	% Including figure files
\usepackage{amsmath}	% Advanced maths commands
\usepackage{subfigure}
\usepackage{float}
\usepackage{natbib} 
\usepackage{hyperref}
\usepackage{tabularx}
\title[The Flaring TOIs]{The Flaring TESS Objects of Interest: Flare Rates for all Two Minute Cadence TESS Planet Candidates}

\author[Ward S. Howard]{
Ward S. Howard,$^{1}$\thanks{E-mail: Ward.Howard@colorado.edu (WSH)}
\\
$^{1}$ Department of Astrophysical and Planetary Sciences, University of Colorado, 2000 Colorado Avenue, Boulder, CO 80309, USA \\
}

\date{Accepted 2022 Mar 2. Received 2022 Mar 1; in original form 2022 Feb 14}

\pubyear{2022}

\begin{document}
\label{firstpage}
\pagerange{\pageref{firstpage}--\pageref{lastpage}}
\maketitle

\begin{abstract}
Although more than 5000 TESS Objects of Interest have been cataloged, no comprehensive survey of the flare rates of their host stars exists. We perform the first flare survey of all 2250 non-retired TOIs with 2 min cadence light curves to measure or place upper limits on their flare rates. We find 93 candidates orbit flare stars and measure their flare frequency distributions. Across the sample, TOIs of $\leq$1.5$R_\oplus$ orbit flare stars more frequently than do TOIs of 1.5$<$R$<$2.75$R_\oplus$, 2.75$<$R$<$4$R_\oplus$, or R$\geq$4$R_\oplus$. We sort all TOI host stars by their flare rate/upper limit, stellar mass, and distance to create a flare ranking metric (FRM) to determine suitability for follow-up. The FRM of each TOI is then checked against the expected signal-to-noise of atmospheric features in transmission spectroscopy to locate the most promising targets. We find 1/4 of terrestrial M-dwarf planets amenable to transmission spectroscopy orbit flare stars. However, none of the M-dwarf hosts to terrestrial planets are currently flaring at sufficient levels for $>$99.9\% atmospheric ozone depletion. We give the first upper limits on the flare rate of the host star to TOI 700 d and explore the flare rates incident on young planets such as DS Tuc Ab.
\end{abstract}


%\interfootnotelinepenalty=10000

\section{Introduction}
The Transiting Exoplanet Survey Satellite (TESS; \citealt{Ricker_TESS}) has vastly increased the number of planets available for detailed characterization; more than 5000 TESS Objects of Interest (TOI; \citealt{Guerrero2021}) have now been detected \footnote{exofop.ipac.caltech.edu/tess}. TOIs are candidate or confirmed transiting planets. With the launch of JWST and increasing numbers of space and ground based measurements of exoplanet atmospheres (e.g. \citealt{Benneke2019,DiamondLowe2020}), exoplanet science is moving beyond discovery into detailed characterization. Exoplanet characterization efforts require a careful treatment of starspots and flares to avoid false positives in atmospheric line or bio-signature detections \citep{Barclay2021, Luger2015} and to probe non-equilibrium atmospheric states resulting from high rates of flaring \citep{Chen2021}.
\par Stellar flares occur when magnetic reconnection in the stellar atmosphere accelerates charged particles into the photosphere, heating the plasma and releasing energy across the electromagnetic spectrum. Low mass stars can remain active for $\sim$10 Gyr \citep{France2020}, frequently emitting superflares 10-1000$\times$ larger than flares from our Sun \citep{Howard2018}. Superflares may impinge on the habitability of terrestrial planets orbiting in the liquid-water habitable zones (HZ) of these stars \citep{tarter2007}, although it remains uncertain if they are a net positive for surface life. A high flare rate may photo-dissociate atmospheric volatiles such as ozone, allowing lethal doses of UV surface radiation \citep{Tilley2019}. On the other hand, the UV energy needed for pre-biotic chemistry on planets orbiting low mass stars can only come from flares \citep{Ranjan2017}.
\par Although flares impact both exoplanet characterization and habitability, no comprehensive survey of the flare rates of the TOI catalog exists. The impact of flaring on each confirmed TOI has been assessed largely on a case-by-case basis. Usually, discovery papers acknowledge whether flares were observed by TESS or not but do not measure the flare frequency distribution or place upper limits on the flare rate. As a result, $\sim$15 TOI host stars have measured flare rates (e.g. \citealt{Gunter2020, Pope2021, Gilbert2021, Bogner2021, Colombo2022arXiv}). Most notably, \citet{Medina2020} performed a flare search on all single low mass stars in the southern sky within 15 pc. By employing injection and recovery testing, they measured flare rates or placed upper limits for 5 of the most promising TOIs for future characterization.
\par
\begin{figure*}
	\centering
	{
		\includegraphics[trim= 0 0 0 0, width=0.92\textwidth]{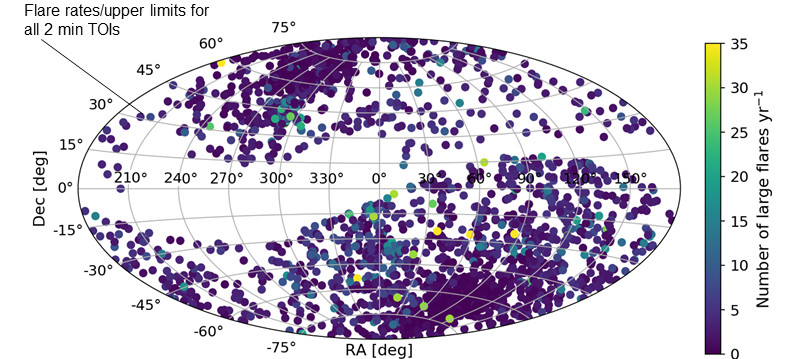}
	}
	%\vspace{-0.2cm}
	\caption{Flare rates or upper limits of all 2 min cadence non-retired TOIs. Upper limits are obtained with flare injection and recovery testing. The colorbar gives the number of large flares yr$^{-1}$, defined here as the rate of flares with an ED$\geq$100 sec. We use ED instead of energy in the color scaling to visually account for the differing contrasts of the same size flare for low mass and high mass stars.}
	\label{fig:flaring_cartograph}
\end{figure*}

 Most of the sky has now been observed continuously for at least 2 months during the TESS Prime and Extended Mission, sufficient to place strong upper limits on the flare rates of low mass TOIs. We carry out the first comprehensive search for flares for all 2250 non-retired TOIs with 2 min light curves (Fig. \ref{fig:flaring_cartograph}). Non-retired TOIs are those that have not been classified as false positives by the TESS mission. Following \citet{Medina2020}, we use injection testing to obtain upper limits on the flare rates of all non-flaring TOIs in order to determine ``worst case scenarios" for atmospheric characterization.

\section{Building the TOI Flare Catalog}\label{TOIsample}
TESS searches the entire sky for both transiting exoplanets and astrophysical variability in a tiling-based approach. The sky is split into 24$\times$96 deg$^{2}$ sectors, which are each observed for 28 days at a time with four on-board 10.5 cm telescopes in a red (600-1000 nm) bandpass. A given sector is therefore revisited with a $\sim2$ year cadence, although overlaps between sectors can increase coverage. Targets in the Continuous Viewing Zone are observed with almost unbroken coverage for a year. Since 2018, TESS has identified more than 5000 exoplanet candidates, or TOIs. We select all TOIs with 2 min photometry for our flare analysis, except for those TOIs with a TESS disposition indicating known false positives (FP). TOIs from 10-30 min cadence light curves are excluded due to a need to resolve flares. In sum, we download Simple Aperture Photometry (SAP) light curves for 2250 TOI planet candidates and 2096 host stars meeting these criteria from the Mikulski Archive for Space Telescopes (MAST).
\par
\begin{figure*}
	\centering
	{
		\includegraphics[trim= 0 0 0 0, width=0.99\textwidth]{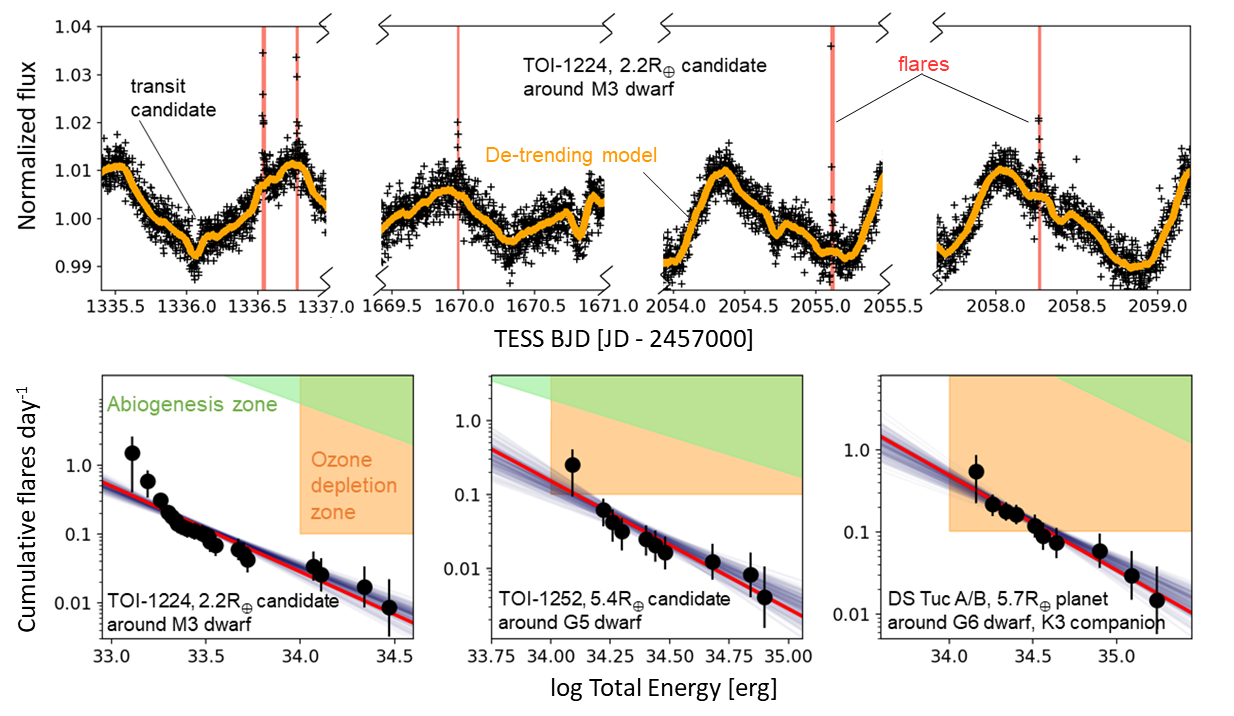}
	}
	\vspace{-0.3cm}
	\caption{Top: Example TOI light curve illustrating the fitting of non-flare variability, flare detections, and the sometimes close occurrence in time of flares and transits. Bottom: Example FFDs of TOIs illustrating the cumulative flare rates and uncertainties of each individual energy y-value, the best fit in red, and fit errors in navy. The \citet{Tilley2019} ozone depletion and abiogenesis regions of \citet{Rimmer2018, Ducrot2020} are over-plotted.}
	\label{fig:methods_overview_fig}
\end{figure*}

 Long-term variability is carefully removed from each light curve before searching for flares. First, the period of variability is estimated from a Lomb-Scargle periodogram of the light curve. The highest peak with a signal/noise ratio (S/N) above 50 is selected as the best period; otherwise the period is set to $\sim$2 d to ensure any out-of-flare variability at longer timescales is removed. Not all long-term variability is strictly periodic. Next, we fit and remove the out-of-flare variability using a Savitzky–Golay (SG) filter with a window size determined by the estimated variability period as described in \citet{Howard_MacGregor2021}. An example of this fitting process is shown in Fig. \ref{fig:methods_overview_fig}. Light curves are then converted into fractional flux units as in \citet{hawley2014}.
\par Flares are identified by an automated algorithm described in \citet{Howard_MacGregor2021} as $\geq$4.5$\sigma$ excursions above the photometric scatter in the detrended light curve. Flare start and stop times are given as the first and last timestamps with fluxes more than 1$\sigma$ above the noise and are subsequently adjusted by eye. Candidates are inspected for FPs. The equivalent duration (ED) of each flare is obtained by integrating the fractional flux between the start and stop times in seconds. The quiescent luminosity of each star $Q_0$ is computed in $T$ in erg $s^{-1}$ using the $T$=0 to flux calibration \citep{Sullivan2015}, the stellar distance, and the $T$ mag using values primarily from the TESS Input Catalog \citep{Bailer_Jones2018, Stassun2019}. Bolometric flare energies are computed as $E_\mathrm{bol} = Q_0\times$ED$\times f_\mathrm{bol}^{-1}$ assuming a canonical $\sim$9000 K flare with a $T$-$E_\mathrm{bol}$ conversion factor of $f_\mathrm{bol}$=0.19 \citep{Howard_MacGregor2021}. In all, we observe 697 flares from 87 unique host stars to 93 TOIs.

\subsection{Flare rates for each TOI}\label{tess_FFDs}
Flare emission from the sun and other stars is described by a power law in which higher energy flares are emitted less frequently than lower energy flares. Flare frequency distributions (FFDs) are computed for each TOI given the cumulative rate at which flares of energy $E$ or greater are observed per day and the total observing time. FFDs are fit in log-log space in the usual way:
\begin{equation}
    \log{\nu}=(1-\alpha)\log{E} + \beta
\end{equation}
Here, $\nu$ is the number of flares with an energy greater than or equal to $E$ erg per day, 1-$\alpha$ describes the frequency at which flares of various energies occur, and $\beta$ determines the overall rate of flaring. We calculate the uncertainty in the cumulative occurrence rate for each energy value with a binomial 1$\sigma$ confidence interval (CI). Rates and uncertainties of the smallest flares are corrected/propagated when below the 100\% completeness fraction as computed below. Previous studies find $\alpha\approx 2$ (e.g. \citealt{Ilin2021a, Medina2020}). We use this information to constrain 1-$\alpha$ when fitting FFDs to stars with less than 5 flares. There is some debate in the literature whether $\alpha=2$ \citep{Medina2020} or depends on factors such as stellar rotation \citep{Seligman2022}. We therefore fit 200 trials for each FFD as shown in Fig. \ref{fig:methods_overview_fig}, varying the y-axis values within their uncertainties and allowing fits across the 1.3$<\alpha<$2.3 literature range (e.g. \citealt{Howard2019,Ilin2021a,Feinstein2022}).

Injection and recovery tests are performed for both flaring and non-flaring TOIs to determine the completeness fraction of flares recovered from each TOI as a function of flare energy following \citet{Glazier2020}. Batches of 100 flares of a given size are randomly injected into separate instances of each light curve and the number of flares recovered is recorded. The process is repeated for increasingly large sets of flares in order to determine when the recovery fraction converges to one. The 10\% completeness fractions (i.e. a proxy for the minimum fraction with sufficient signal to correct for incompleteness), and 100\% completeness levels are reported in Table \ref{table:all_toi_tab}, respectively. Injection testing of the non-flaring TOIs enables upper limits to be placed on their flare rates when combined with the total observation time. First, the upper 1$\sigma$ binomial uncertainty on the number of flares of energy $E_\mathrm{min}$ observed across the total observation time in days is measured. Then the rate of flares per day of $\geq E_\mathrm{min}$ is fit with 200 FFD trials across the full $\alpha$ range. When no flares are observed, we report the number of flares as zero and the FFD fit parameters for the upper limit in Table \ref{table:all_toi_tab}.
\begin{table*}
\renewcommand{\arraystretch}{1.6}
\caption{Catalog of all 2250 TOIs with 2 min cadence TESS data and a non-FP TESS disposition}
\begin{tabular}{p{0.7cm} p{0.5cm} p{0.5cm} p{0.5cm} p{0.6cm} p{1.3cm} p{0.7cm} p{0.7cm} p{0.6cm} p{0.9cm} p{1.1cm} p{0.7cm} p{0.7cm} p{1.3cm}}
\hline
TOI & R$_p$ & TSM & FRM & TFOP & TIC-ID & M$_*$ & $d$ & Bin. & N$_{fl}$ & log $E_\mathrm{100}$ & 1-$\alpha$ & $\beta$ & N$_{super}$ \\
 & [$R_\oplus$] &  &  &  &  & [$M_\odot$] & [pc] & [bool] &  & [erg] &  &  & [fl. yr$^{-1}$] \\
\hline
134.01 & 1.74 & 135 & 0.94 & Y & 234994474 & 0.59 & 25.1 & 0 & 0 & 33.44 & -0.80 & 25.2 & 3.0 \\
177.01 & 2.13 & 150.0 & 0.932 & Y & 262530407 & 0.52 & 22.5 & 0 & 2 & 33.32 & -0.80 & 25.18 & 4.2 \\
218.01 & 1.83 & 133.0 & 0.980 & Y & 32090583 & 0.26 & 52.3 & 0 & 152 & 33.27 & -1.67 & 54.84 & 4.4 \\
540.01 & 0.97 & 38.0 & 0.998 & Y & 200322593 & 0.17 & 14.0 & 0 & 14 & 32.45 & -1.66 & 52.9 & 0.07 \\
%  &  &  &  &  &  &  & ... &  &  &  &  &  &  \\
\hline
\end{tabular}
\label{table:all_toi_tab}
{\newline\newline \textbf{Notes.} A subset of the flaring parameters of 2250 non-retired TOIs observed at 2 min cadence. The full table is available in machine-readable form. Columns shown are TOI, planet radius, TSM, FRM, TFOP disposition, TIC ID, stellar mass, stellar distance, a flag whether binarity in Gaia DR2 \citep{Kervella2019} or source contamination is likely present, number of flares observed, smallest energy flare that would be recovered 100\% of the time, the FFD fit parameters 1-$\alpha$, $\beta$, and the rate of 10$^{34}$ erg superflares. Other columns in the machine readable table not shown here include orbital distance, orbital period, coordinates, Gaia ID, observation time, rate of ED=100 sec flares yr$^{-1}$, the energy and ED of 10\% completeness limits, and uncertainties where applicable.}
\end{table*}

\section{Suitability for atmospheric characterization and flare rates}\label{sorted_by_flaring}
TOIs are ranked by flaring on their suitability for follow-up, where the most promising targets for follow-up are low mass stars with low flare rates. Each TOI is assigned a ranking between zero and one, with one being highly-favorable for follow-up and zero very unfavorable. This TOI flare ranking metric (FRM) sorts TOIs by increasing flare rate within distinct groupings organized by stellar mass and distance, with nearby, low mass stars ranked higher than distant, high mass stars. The quiescent brightness of higher mass stars reduces the contrast of a given size flare and induces a flare detection bias toward higher energies from larger stars. Grouping the TOIs by stellar mass minimizes this bias and facilitates comparisons of the relative suitability for follow-up between stars of different masses. Grouping by distance accounts for the increased priority for follow-up of stars in the solar neighborhood. The first several groups are ordered as follows: 0.1-0.3$M_\odot$ late M-dwarfs within 15 pc, 0.3-0.6$M_\odot$ early M-dwarfs within 15 pc, 0.1-0.3$M_\odot$ late M-dwarfs beyond 15 pc, 0.3-0.6$M_\odot$ early M-dwarfs beyond 15 pc, 0.6-0.75$M_\odot$ late K-dwarfs within 36 pc, 0.75-0.9$M_\odot$ early K-dwarfs within 36 pc, 0.6-0.75$M_\odot$ late K-dwarfs beyond 36 pc, 0.75-0.9$M_\odot$ early K-dwarfs beyond 36 pc, and so on.

The grouping order continues to higher-mass stars until all TOIs are included. \citet{Winters2021} gives 15 pc as the demarcation distance between nearby and distant M-dwarfs. We compute an equivalent demarcation distance for the larger mass bins such that the apparent magnitude of stars in each bin is approximately comparable to M-dwarfs observed from 15 pc.
\par Sorting by flare rates/upper limits within each of these mass-distance groupings broadly agrees with our intuitions that a nearby rocky planet orbiting an M-dwarf flare star is less suitable for characterization than a a nearby rocky planet orbiting a non-flaring M-dwarf, while still having a higher priority for follow-up than a planet orbiting a distant G-dwarf from which no flares were observed. The catalog ranking for the top 10\% of targets is shown in Fig. \ref{fig:results_fig} to illustrate the effects of flare rate, stellar mass, and distance on the flare ranking. Specific use cases for our catalog may benefit from sorting the flare rates of the TOIs with different criteria or in a different order, which may be done using the columns provided in the machine-readable version of Table \ref{table:all_toi_tab}. The final flare ranking is generated by splitting the catalog into three categories based on TFOP working group disposition in ExoFOP-TESS. TFOP-confirmed planets and active candidates are placed first, then ambiguous candidates second, and candidates retired as FPs third. Within each TFOP category, the mass, distance, and flare rate ordering is preserved.
\par Once each TOI is assigned its final flare ranking, we cross-correlate the flare catalog with those targets for which a 10 hr JWST program could detect exoplanet atmospheres. Following \citet{Kempton2018}, we compute a transmission spectroscopy metric (TSM) for each TOI in our sample as described by Equation 1 of \citet{Kempton2018}, reproduced below:
\begin{equation}
    \mathrm{TSM} = f_s \times \frac{{R_p}^3 T_\mathrm{eq}}{M_p {R_{*}}^2} \times 10^{-m_J/5}
\end{equation}
Here, $f_s$ is the scale factor as computed in Table 1 of \citet{Kempton2018}. $R_p$ is the ExoFOP planet radius, $T_\mathrm{eq}$ is the estimated planet equilibrium temperature, $M_p$ is the estimated planet mass, $R_*$ is the estimated stellar radius \citep{Torres2010, Mann2015}, and $m_J$ is the $J$ magnitude estimated from \citet{Kraus2007}. We refer the reader to \citet{Kempton2018} for details. The TSM scales with the signal-to-noise of exoplanet atmospheric features, assuming a cloud-free atmosphere and 10 hr of JWST observations. \citet{Kempton2018} define a detectable atmosphere to have TSM$>$10 for terrestrial planets and TSM$>$90 for larger planets. Since our inputs to the TSM equation are estimated from ExoFOP, we caution that terrestrial planets just above or below TSM=10 may be consistent with slightly lower or higher values across the cutoff. Highly-ranked TOIs in our flare catalog that also have high TSM values are particularly compelling for follow-up.

\section{Key results of the TOI flare catalog for atmospheric follow-up}\label{key_results_sect}
Across the sample, we find small ($<$1.5$R_\oplus$) TOIs are more likely to orbit flare stars than larger planets as shown in Fig. \ref{fig:results_fig}. This is a result of the high signal to noise of small planets orbiting low mass stars and the oftentimes high flare rates of these stars.

\subsection{Terrestrial planets potentially suitable for follow-up}\label{some_exciting_targets}
M-dwarf planets or likely candidates of $<$1.5$R_\oplus$ flagged as suitable for JWST transmission spectroscopy make up 35 out of 2250 TOIs in the catalog. Of these, nine or 25.7$\substack{+12 \\ -7}$\% orbit stars observed to flare. These TOIs are LHS 3844 b, the L 98-59 system, the LTT 1445 system, TOI 486.01, TOI 540 b, the GJ 1132 system, TOI 1450.01, and TOI 2267.01 as shown in Fig. \ref{fig:results_fig}. Three orbit flare stars with FFDs previously reported from \citet{Medina2020}. The flare rate of LTT 1445 was previously measured by \citet{Howard2019} but the flares likely come from the BC component \citep{Winters2019}. \citet{Bogner2021} fits dozens of FFDs which includes the host stars to LHS 1815 b, the L 98-59 system, TOI 486, and TOI 1450 but does not report FFD fit parameters for these stars.
\par We do not detect flares from the remaining 26 terrestrial TOIs amenable to characterization, but place upper limits on their flare rates in Table \ref{table:all_toi_tab}. More than 90\% of these did not have previous upper limits. Each of the 35 flagged TOIs is in the top 10\% of our flare suitability ranking, and even the most active do not reach the ozone depletion zone of \citet{Tilley2019}, although some disequilibrium chemistry is consistent with the rates/upper limits of several TOIs \citep{Howard2018}.

\subsection{Other flaring and non-flaring TOIs of note}\label{other_exciting_targets}
\begin{figure*}
	\centering
	{
		\includegraphics[trim= 0 0 0 0, width=0.99\textwidth]{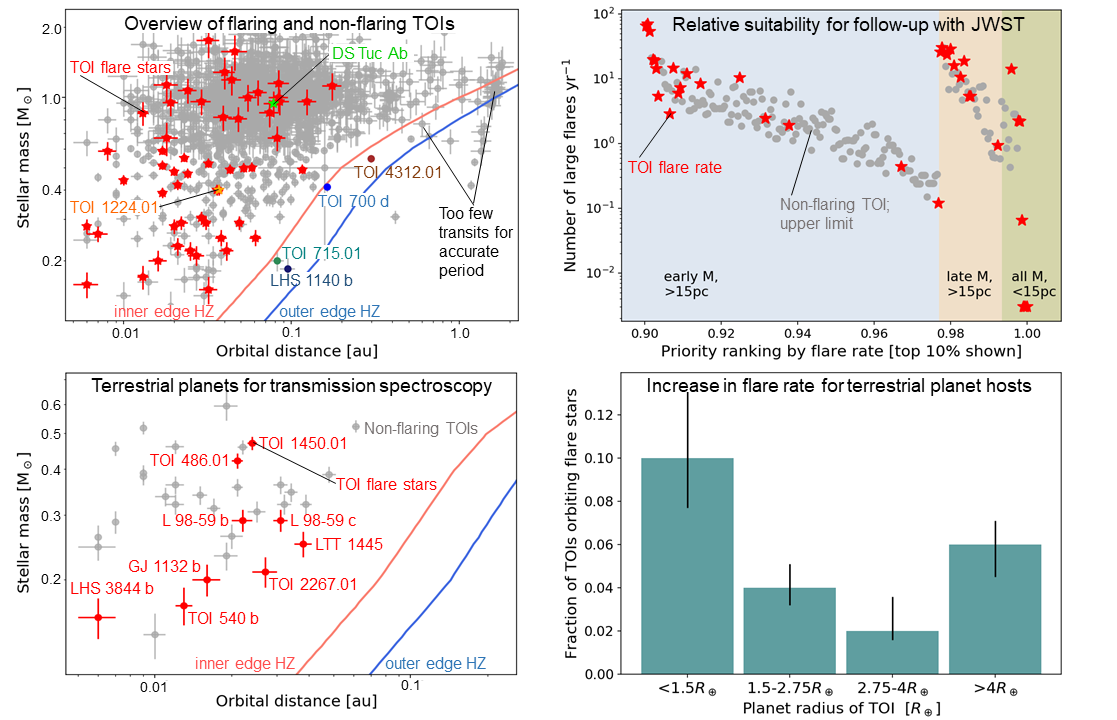}
	}
	%\vspace{-0.2cm}
	\caption{Top left: All non-retired TOIs orbiting flaring (red) and non-flaring (grey) host stars of $\leq$2M$_\odot$. TOI 1224, DS Tuc Ab \citep{Newton2019_DSTuc}, and TOIs in the HZ of \citep{Kopparapu2013} are noted. Top right: The flare ranking metric of \S \ref{sorted_by_flaring}. Bottom left: All non-retired TOIs around M-dwarfs amenable to transmission spectroscopy, with the same color scheme as above. Note the HZ planets from the top left panel fail the \citet{Kempton2018} TSM criteria for a 10 hr JWST program. Bottom right: smaller TOIs orbit flare stars more frequently than larger TOIs, with 1$\sigma$ binomial CIs.}
	\label{fig:results_fig}
\end{figure*}

The flare rates of the remaining TOIs in our sample enable characterization of super-earths, mini-Neptunes, and several planets in the HZ. Terrestrial HZ planets in our sample include TOI 700 d, LHS 1140 b, TOI 715.01, and TOI 4312.01 (Fig. \ref{fig:results_fig}). The first two are among the most exciting planets for large follow-up programs \citep{Gilbert2020_TOI, Wunderlich2021}. We report upper limits on the rate of 10$^{34}$ erg superflare yr$^{-1}$ of 0.25 and 0.71 for these TOIs, respectively. TOI 715.01 is a 1.7$R_\oplus$ super-earth orbiting a likely M-dwarf. It emits no more than 0.15 10$^{34}$ erg superflare yr$^{-1}$. TOI 4312.01 is a 2.4$R_\oplus$ candidate with an upper limit of 4.7 10$^{34}$ erg superflare yr$^{-1}$.

\section{Conclusions}\label{conclude}
We present the first comprehensive TOI flare survey. A new flare ranking metric is created to aid proposers in selecting planets for further characterization. The metric of every TOI is cross-matched against a S/N metric for a 10 hr JWST program. As a result, we discover that $\leq$1.5$R_\oplus$ TOIs orbit flare stars more frequently than do larger planets, and find 1/4 of all $<$1.5$R_\oplus$ TOIs around M-dwarfs accessible to transmission spectroscopy with JWST orbit flare stars. Even though none of the hosts of terrestrial planets suitable for characterization reach the flares rates needed for complete ozone loss, it has been shown that flare rates a factor of $\sim$10$\times$ lower may still induce significant dis-equilibrium states \citep{Howard2018}. These lower rates are comparable to the flare rates and upper limits of the TOIs presented here. Therefore, spectral modeling of dis-equilibrium chemistry and atmospheric survival will be needed on a case-by-case basis and population level to understand atmospheric signals. In addition to JWST, the upcoming 2029 launch of the Ariel mission to characterize the atmospheric properties of exoplanets \citep{Tinetti2013_Ariel} would benefit from our catalog. Finally, our sample of flaring TOIs provides an excellent dataset for optical counterparts to radio star-planet interactions \citep{Pope2021}.  

\section*{Acknowledgements}\label{acknowledge}
We would like to thank the anonymous referee for improving the work. WH would like to thank Meredith MacGregor for helpful conversations on TOI flaring. WH acknowledges funding support by the TESS Guest Investigator Program GO 3174. This paper includes data collected by the TESS mission. Funding for the TESS mission is provided by the NASA Explorer Program. This research has made use of the Exoplanet Follow-up Observation Program website, which is operated by the California Institute of Technology, under contract with the National Aeronautics and Space Administration under the Exoplanet Exploration Program. This work has made use of data from the European Space Agency (ESA) mission {\it Gaia} (\url{https://www.cosmos.esa.int/gaia}), processed by the {\it Gaia} Data Processing and Analysis Consortium (DPAC, \url{https://www.cosmos.esa.int/web/gaia/dpac/consortium}). Funding for the DPAC has been provided by national institutions, in particular the institutions participating in the {\it Gaia} Multilateral Agreement.

{\it Facilities: TESS}

\section*{Data availability}\label{availdata}
The data underlying this article are available in the MAST archive at Space Telescope Science Institute, at https://doi.org/10.17909/t9-nmc8-f686

\bibliographystyle{mnras}
\bibliography{paper_references}

% Don't change these lines
\bsp	% typesetting comment
\label{lastpage}
\end{document}